\documentclass[superscriptaddress,aps,prb,floatfix,footinbib]{revtex4-1}
\usepackage{graphicx}
\usepackage{bm}
\usepackage{hyperref}
\usepackage{amsmath}
\usepackage{}

\newlength{\figwidth} 
\newcommand{\beq}{\begin{equation}}

\newcommand{\beql}[1]{\begin{equation}\label{#1}}
\newcommand{\eeq}{\end{equation}}
\newcommand{\bsp}{\begin{split}}
\newcommand{\esp}{\end{split}}

\newcommand{\Eq}[1]{Eq.~(\ref{#1})}

\newcommand{\Fig}[1]{Fig.~\ref{#1}}

\begin{document}
\title{Energy transfer channels of the plasmon excitation process in STM tunnel junctions}
\author{Yaoqin Lu}
\affiliation{School of Optical and Electronic Information, Huazhong University of Science and Technology, Wuhan 430074,  China}
\author{Yuntian Chen}\email{yuntian@hust.edu.cn} 
\affiliation{School of Optical and Electronic Information, Huazhong University of Science and Technology, Wuhan 430074,  China}
\affiliation{Wuhan National Laboratory of Optoelectronics, Huazhong University of Science and Technology, Wuhan 430074,   China}

\author{Jing Xu}
\affiliation{School of Optical and Electronic Information, Huazhong University of Science and Technology, Wuhan 430074,  China}
\affiliation{Wuhan National Laboratory of Optoelectronics, Huazhong University of Science and Technology, Wuhan 430074,   China}
\author{Tao Wang}
\affiliation{Institute of Materials Research and Engineering, A*STAR (Agency for Science, Technology and Research), 2 Fusionopolis Way, 08-03 Innovis, Singapore 138634, Singapore}
\author{Jing-Tao L\"u}\email{jtlu@hust.edu.cn} 
\affiliation{School of Physics, Huazhong University of Science and Technology, Wuhan 430074, P. R. China}

\begin{abstract}
We study the decay of gap plasmons localized between a scanning tunneling microscope tip and metal substrate, excited by inelastic tunneling electrons. The overall excited energy from the tunneling electrons is divided into two categories in the form of resistive dissipation and electromagnetic radiation, which together can further be separated into four different channels, including SPP channel on the tip, SPP channel on the substrate, air mode channel and direct quenching channel. We find that most of the excited energy goes to surface plasmon polaritons on the metallic STM tip, rather than that on the substrate. The direct quenching in the apex of tip also takes a considerable portion especially in high frequency region. 
\end{abstract}
\maketitle

\section{Introduction}
Since its invention in 1981, the scanning tunneling microscopy (STM) has become one of the most powerful tools to characterize nanostructures on metal surface\cite{Binnig1981,Binnig1982}. It uses the quantum tunneling of electrons from the tip to the metal surface to `measure' the local density of states of the metal surface or adsorbate on the surface. Meanwhile, the nano-gap between the tip and surface hosts localized plasmon modes (gap modes) that can be used to localize the electromagnetic field in nanoscale regime. Gimzewski \emph{et al}. observed for the first time that, the gap modes can be excited by tunneling electrons at high enough bias\cite{Gimzewski1988}. Radiative decay of these localized modes gave rise to light emission that was detected from the far field at the same side of the STM tip, coined as STM induced luminescence (STML). Afterwards, the effect of tip, surface shape, gap size, dielectric environment, types of metal on the light emission properties are investigated \cite{Berndt1991,Berndt1993,Berndt1993-2,Berndt1998,Chen2009,Yu2016,Chen2014,Nilius2000,Ushioda2000}. The combination of plasmonic and molecular luminescence is also studied \cite{Qiu2003,Dong2004,Schneider2012,Lutz2013,Imada2016,Zhang2016,Zhang2017,Imada2017}. Recently, the change of STML properties as a function of tip-surface distance is investigated from the tunneling to contact regime both for metal and molecular junctions\cite{Schull2009,Schneider2010,Schneider2012}. The relation between optical yields and finite frequency shot-noise of electrons is revealed \cite{Schull2009,Schneider2010,Schneider2012,Lu2013,Xu2015,Kaasbjerg2015}.

Theoretical analysis of the light emission efficiency reveals that the gap plasmons are mainly excited by inelastic tunneling electrons in the gap, instead of hot luminescence in the electrodes \cite{Persson1992}. Electromagnetic simulations are also conducted to study the emission spectrum. In the early theoretical study of STML, metallic sphere is used to approximate the tip for the calculation of the plasmon emission spectrum in the far field\cite{Rendell1981,Johansson1990,Johansson1998}. The dependence of emission spectrum on the material, sphere radius, sphere-substrate distance and applied voltage was systematically studied. Subsequently, J. Aizpurua \emph{et al.} adopted a hyperbolic tip geometry which is similar to the tip used in experiment.
A more precise relationship between the tip shape and the spectrum is obtained\cite{Aizpurua2000}.  Recently, it was found that the gap mode can also excite the surface plasmon polaritons (SPPs) propagating along the metal surface\cite{Novotny2011,Wang2011}. This provides an efficient, local, electrical way of launching SPPs in optical structures, and its applications received considerable attention recently\cite{Wang2014,Cao2014,Wang2015,Du2016,Du2017,Cazier2016}. These results suggest that the electrical excited gap plasmon modes have several optical decay channels. 

A natural question to ask is how much of the energy is transferred into different decay channels. This is important for engineering the energy separation among different channels\cite{Greffet2016} and for improving the optical yields in STML. As far as we know, it has not received enough attention yet. In this work, we try to answer this question from classical electromagnetic simulation. 

The paper is organized as follows: In Section 2, we present the  geometric setting for STML experiments and build the light emission model based on the point dipole approximation of the tunnel junction, and further classify the emission channels with particular emphasis on separating the direct quenching from the total resistive dissipation, which is usually difficult from perspective of classical electromagnetism. In Section 3, the numerical results are given and analyzed, showing that the major emission is distributed into the SPPs on the tip. Finally, Section 4 concludes the paper.

\section{Geometry structure and emission channels of STM luminescence}

\begin{figure}[htb!]\centering
\includegraphics[scale=0.14]{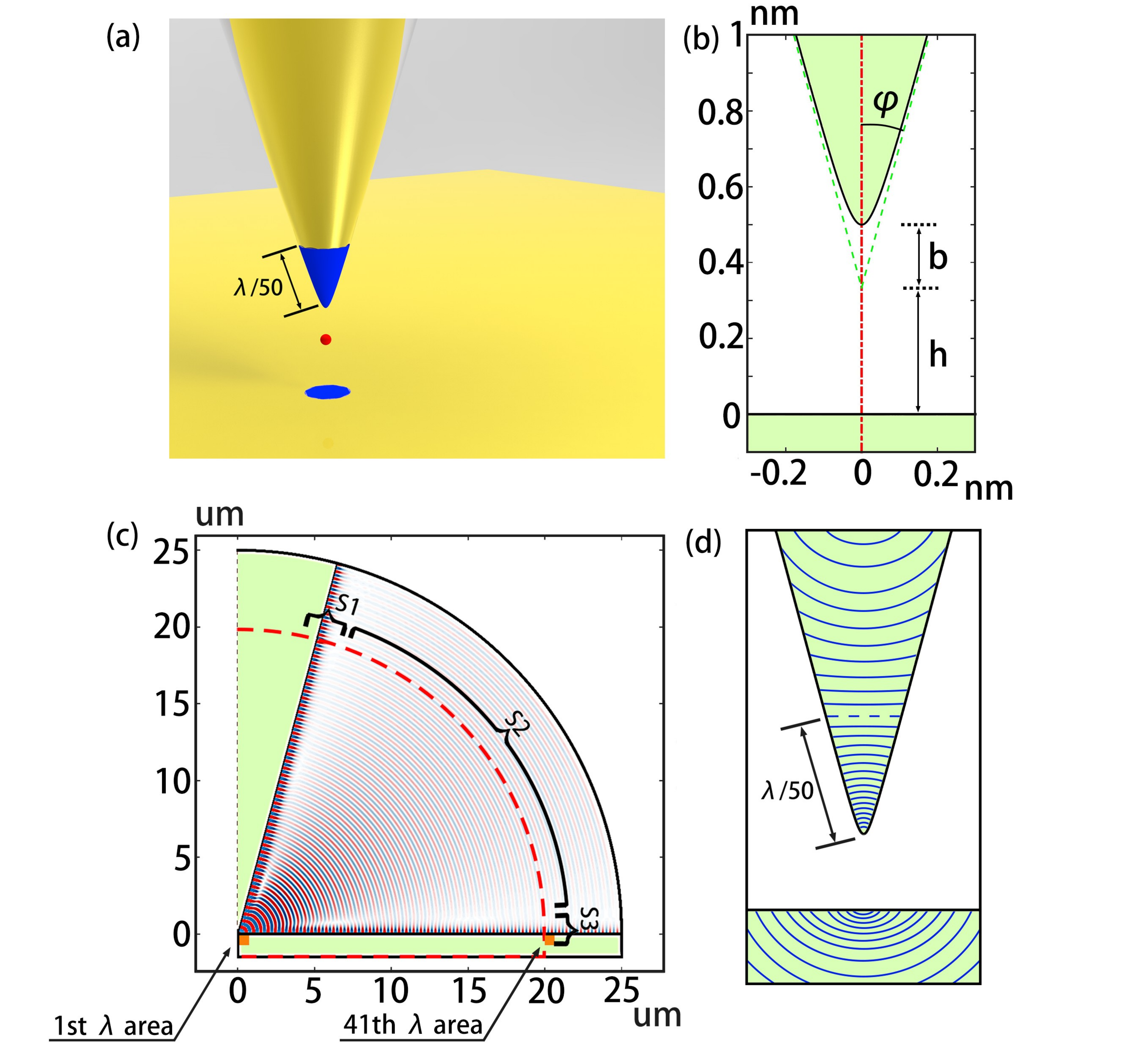}
\caption{ (a) A three dimensional model of STM structure. Blue areas represent the direct quenching, which in the apex shows as a cone, the generatrix is $\lambda$/50. (b) The sketch of the silver tip apex, it is depicted by the hyperbolic geometry, $h+b$ represents the gap distance, the value of $b$ determines the acuity of the apex. (c) The 2D plane of the STM structure and the radiative modes pattern. S1 and S3 are used to collect the energy of SPPs, while S2 is used to collect the energy of air mode. The orange  blocks  located at $x=0$ ($x=20$) $\mu$m along x-axis, which are marked as "1st (41th) $\lambda$ area", are used to calculate the resistive dissipation of SPPs propagation losses in the substrate. (d) The contour plot of resistive dissipation is sketched. The direct quenching contours are concentric to the apex. In contrast, the contours that are much far away from the apex tend to bend towards the tip surface, indicating the dominating effect from the  propagation losses of SPPs. The resistive dissipation in the tip can be separated by the dash line but that in the substrate can't be distinguished clearly.} 
\label{fig1}
\end{figure}

\Fig{fig1} (a) shows the structure considered in this work, with a metallic tip positioned right above a metallic surface. The geometric properties of the tip is depicted in \Fig{fig1} (b). Although some efficient methods are proposed in the literature to take into account the quantum effects\cite{RubenEsteban2011,Zhuwenqi2016}, we follow a classical electrodynamics simulation here. The energy source is approximated by an electrical dipole located between the tip and surface (red dot in \Fig{fig1} (a)). Considering the rotation symmetry of the structure, a two-dimension (2D) rotational model is adopted, and the electric dipole is implemented as a magnetic current loop with tiny radius, i.e., 0.1nm, in comparison  with the work wavelength with range from 400 nm to 900 nm.

\subsection{Nanophotonic structure to enhance STM luminescence}
A snapshot of the emitted electromagnetic radiation in the reduced 2D plane is presented in \Fig{fig1} (c). We have used the parameters of silver\cite{Johnson1972} in the simulation. The cone angle of the tip is 15 degrees. The shape of the apex also has significant impact on the spectrum. We use the hyperbolic tip geometry (${\frac{{{{\left( {z - h} \right)}^2}}}{{{b^2}}} - \frac{{{\rho ^2}}}{{{{\left( {b\tan \varphi } \right)}^2}}} = 1}$) \cite{Aizpurua2000} for its relatively good fidelity of representing the tips used in experiment as sketched in \Fig{fig1} (b), where $h+b$ is the distance between the apex and the metal substrate and $b$ is the distance between the intersection of asymptotes and the tip apex. Evidently, the apex becomes flatter as $b$ increases on the premise of fixed $h$. In our model, the ratio of $h$ and $b$ is $\frac{h}{b}=2:1$.

\subsection{Emission channels in STM induced luminescence}
Consider a volume $V$ with its surface $\partial V$, the Poynting theorem states that the time rate of electromagnetic energy change within $V$ plus the net power flowing out of $V$ through $\partial V$ is equal to the negative of total work done on the charges within $V$. Here, we consider electromagnetic waves with  time harmonic oscillations,  thus the time rate of electromagnetic energy change within $V$ vanishes, leading to the reduced energy conversation law, given as follows,
\beq
\label{poynting}
-\int_V {\bm J} \cdot {\bm E} dV =\mathop{{\int\!\!\!\!\!\int}\mkern-21mu \bigcirc}\nolimits_{\partial V} 
 ({\bm E \times \bm H) \cdot d\bm A}.    
\eeq 

In \Eq{poynting}, the total current density term $\bm J$ can be split into two terms ($\bm J={\bm J}_{s}+{\bm J}_{c}$), i.e., the source current density term $\bm J _s$ and the polarization current density term $\bm J_c= \frac{d \left[\left({\epsilon}_r-1\right)\epsilon_0 \bm E\right]}{dt}$. The source current density term measures the total external work done on the electrons to generate the electromagnetic radiation, whereas the polarization current density term measures the dissipation rate into the resistive dissipation due to material losses. Indeed, in our setting the source current density term $\bm J_s$ corresponds to the electron tunneling between the tip and the substrate, i.e., the source of generated electromagnetic energy, while the polarization current density  $\bm J_c$ is the energy sink of the electromagnetic radiation.   

The volume $V$ considered here corresponds to the region by rotating the area marked by red dash line along the y-axis, as shown in \Fig{fig1} (c), which contains the STM tunnel junction. The closed surface $\partial V$ associated with $V$ can be divided into four parts, i.e., $\partial V=S_1+S_2+S_3+S_r$. The net energy flowing through $S_1$ and $S_3$ are the SPP channels propagating along the tip and the substrate respectively, and the photons flowing through $S_2$ are considered as the air mode. $S_r$ is the rest surface of $\partial V$ and none of energy flows through it. Since only the metals have the material losses, the resistive dissipation that absorbs energy from the electromagnetic radiation is exclusively from metals, which can be further separated into direct quenching and the propagation losses of SPPs. The direct quenching occurs locally, as marked as the small blue areas on the tip and the substrate in \Fig{fig1} (a). In contrast, the resistive dissipation from the propagation losses occurs non-locally, as long as the field amplitude of propagating SPP mode is large enough to excite the electron-hole pairs inside the metals. The difference between the two dissipation mechanisms of the electromagnetic radiation will be discussed  and used to separate the direct quenching from the total dissipation. 

There are four emission channels during the  electrons tunneling across the STM junction. The first one is the free propagating photons coined as the air mode channel, which is generated by the tunnel junction and radiated into free space. Secondly, part of power from the oscillating dipole is transfered into resistive dissipation around the local area as indicated by the small blue areas on the tip and the substrate in \Fig{fig1} (a). This is called the direct quenching channel. Thirdly, there are two SPP channels that funnels the electromagnetic radiation along the metal-air interface, which are coined as the tip and substrate SPP channels in the following part of this paper. The tip (substrate) SPPs propagate along the metal-air interface along the surface of the tip (substrate).  Their field amplitudes decay due to the propagation losses, which eventually generates resistive dissipation. In  summary, the direct quenching is indicated by the small blue areas in \Fig{fig1} (a), the air mode is indicated by integrated power flux over the surface area  $S_2$ shown in \Fig{fig1} (c) while the tip (substrate) SPPs energy is the sum of integrated power flux  over $S_1$ ($S_3$) and the propagation losses in the tip (substrate).

\subsection{Extraction of direct quenching from the total resistive dissipation}
In classical electromagnetism, it is not trivial to distinguish direct quenching from the propagation losses of the SPP modes in the vicinity of the tunnel junction, since both of them yield the same type of resistive dissipation, i.e, heating. The quantitative assessment of the light emission that are funneled into different channels need to be acquired. The reason that we numerically measure the integrated Poynting vectors 40 wavelengths away from the tunnel junction (see \Fig{fig1} (c)) is to avoid the overlap of SPPs and air mode. To estimate the original power of SPPs excited by the tunneling electrons directly, we need to add up the resistive dissipation in metal caused by SPPs propagation. However, we have mentioned that in the vicinity of the tunnel junction, there exists direct quenching which is also a kind  of resistive dissipation. Here comes the question how to separate the resistive dissipation caused by inelastic tunneling electrons and SPPs.

In this work, the separation of the direct quenching from the total  resistive dissipation at the apex and the substrate are treated differently. As to the apex, the method is to set up a separation line of the tip, marked as blue dash line in \Fig{fig1} (d). For the direct quenching is extreme large in the tiny part of the apex and decreases dramatically in a further place, we assume it only exists in the area surrounded by the separation line in the apex, outside this area the resistive dissipation is only attributed to SPPs propagation. From the analysis of the resistive dissipation  contours, the direct quenching contours are concentric to the apex, but the SPPs propagation loss contours tend to bend towards the surface of the tip as shown in \Fig{fig1} (d). It means the resistive dissipation caused by the SPPs is dominant in the area away from the tip apex. We find the separation distance shown in \Fig{fig1} (a) can be set to 1/50 of the wavelength, which turns out to a good approximation  in this model.

As to the substrate, the aforementioned method does not apply, since  there is no clear separation line evident from the contour plot of the resistive dissipation shown in \Fig{fig1} (d). The overlap of direct quenching and propagation losses mainly appears in the vicinity of the tunnel junction such as the 1st $\lambda$ area in \Fig{fig1} (c). The extraction of propagation losses in the 1st $\lambda$ area can be achieved by multiplying the propagation losses in the 41th $\lambda$ area (in this area, the resistive dissipation exclusively  originates from  the propagation losses of SPPs and can be easily obtained from our numerical model) by a certain ratio. The ratio can be calculated from the semi-analytical formulation of the cylindrical SPPs provided by S\"ondergaard \cite{Sondergaard2004}  for  a dipole  emitter oriented along $z$-axis. With the expression of electrical field, the resistive dissipation in the 1st/41th $\lambda$ area due to the  propagation losses of SPPs can be calculated as follows,
\beq
\label{Qrh}
Q=\pi \omega\epsilon_0  \int _{0(40\lambda) }^{\lambda(41\lambda)} d\rho \int _{-\lambda}^{0}  dz \rho  \texttt{Im}\left[\epsilon_{r} \left({{E_{\rho}}} E_{\rho}^{*} + {E_{z}} {E_{z}}^{*} \right)\right], 
\eeq
where $\epsilon_{r}$ is the dielectric function of the metallic substrate, and see $E_{\rho}$ and $E_z$ in Appendix A. Based on \Eq{Qrh}, one is able to obtain the ratio between  the propagation losses in the 1st $\lambda$ area and that from the  41th  $\lambda$ area. Since the exact dissipation of the propagation losses in  the 41th  $\lambda$ area is known, one could immediately calculate the restive dissipation from the propagation losses in 1st $\lambda$ area. As such, one could extract the direct quenching by subtracting the resistive dissipation of propagation losses from the total dissipation in the 1st $\lambda$ area.

\section{Results}
\begin{figure}\centering
\includegraphics[scale=0.5]{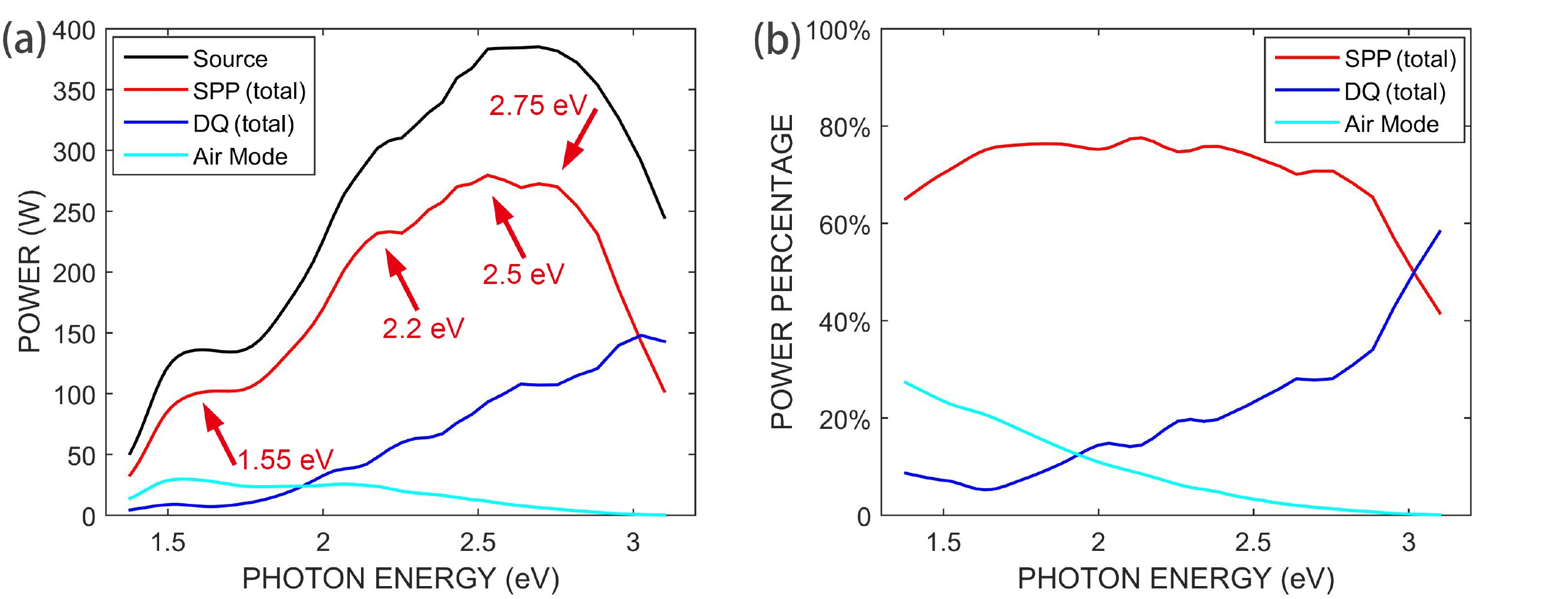}
\caption{ (a) The total power (black) and its distribution into different channels as a function of photon energy. The red, blue and cyan lines represent energy goes into the SPPs, direct quenching and the air mode channels, respectively.  The SPPs and direct quenching (DQ) include both tip and substrate contributions. (b) The percentage of total power into the three kinds of channels as in (a).}
\label{fig2}
\end{figure}

We proceed to discuss the calculated results based on the aforementioned model and to classify the emission energy transfer in the STM junction. The tunneling electrons can be seen as the emitters, modeled by a magnetic current loop.  There are two approaches to obtain the total emitted power according to the Poynting theorem:  one is to integrate the Poynting vector over a closed surface surrounding the emitter, the other is to integrate the tangential component of magnetic field along the circle with the magnetic current. As a self-benchmark, we carry out the two different procedures to obtain the total emitted power. Indeed, the two approaches yield exactly the same emission power. 

The four curves shown in \Fig{fig2} (a) represent the energy radiated by the source, the total power flowing into the sum of two SPP channels, the direct quenching channel, and the air mode channel, respectively. The SPP channels take a dominant part of the source energy. It has a similar lineshape with the total power, sharing the same trends and peaks. The power of SPP channels decreases to zero when the wavelength is less than 1.4 eV or larger than 3.1 eV, such decrease looks more drastic in the high frequency region than in the low frequency region. Such dramatic dropping of emission into SPP channels at high frequencies is due to the cutoff frequency of SPP modes propagating along the interface. The maximum value of the SPP appears at the photon energy of 2.5 eV. In addition, there are some other peaks at the photon energy of 1.55 eV, 2.2 eV, 2.75 eV. The power into the air mode channel is much smaller than the SPP channels. The emission power into  the air mode is almost zero in the high frequency region and increases gradually when the frequency decreases, it increases to the maximum at the photon energy of 1.5 eV. In consistency with the experiment, the air mode is detected in the far field\cite{Schneider2010}, which has approximately  the same  peak position. The direct quenching in this model is also strongly frequency dependent. In the high frequency region, the direct quenching takes up nearly half of the source energy. When the frequency decreases, the direct quenching decreases monotonously. The tremendous quenching is partly resulted from the nanoscale gap distance\cite{Greffet2016}. Comparing the SPP mode and the air mode, we find that the spectral lineshape  of emitted  SPP has a broad peak located at high frequency region, i.e, from blue to green; while  the spectral peak of  the air mode emission lies in-between infrared and red frequencies.

In Fig. 2(b), the fractional contribution of the SPP, the direct quenching and the air mode channel are studied. The SPP channels occupy roughly 70\% of the total emission energy in most of the frequency region. It starts to decrease significantly  when the photon energy is larger than 2.8 eV. On the other hand, the fraction of  air channel decreases smoothly when the frequency increases. This indicates that increasing the source power, i.e., increasing the number of tunneling electrons, can improve the emitted power into the air mode more efficiently  in the low frequency region. The fraction of the direct quenching  increases for larger frequency, which indicates that the adoption of conducting metals with low-impedance may be useful to decrease the directing quenching in the high frequency region.

\begin{figure}\centering
\includegraphics[scale=0.5]{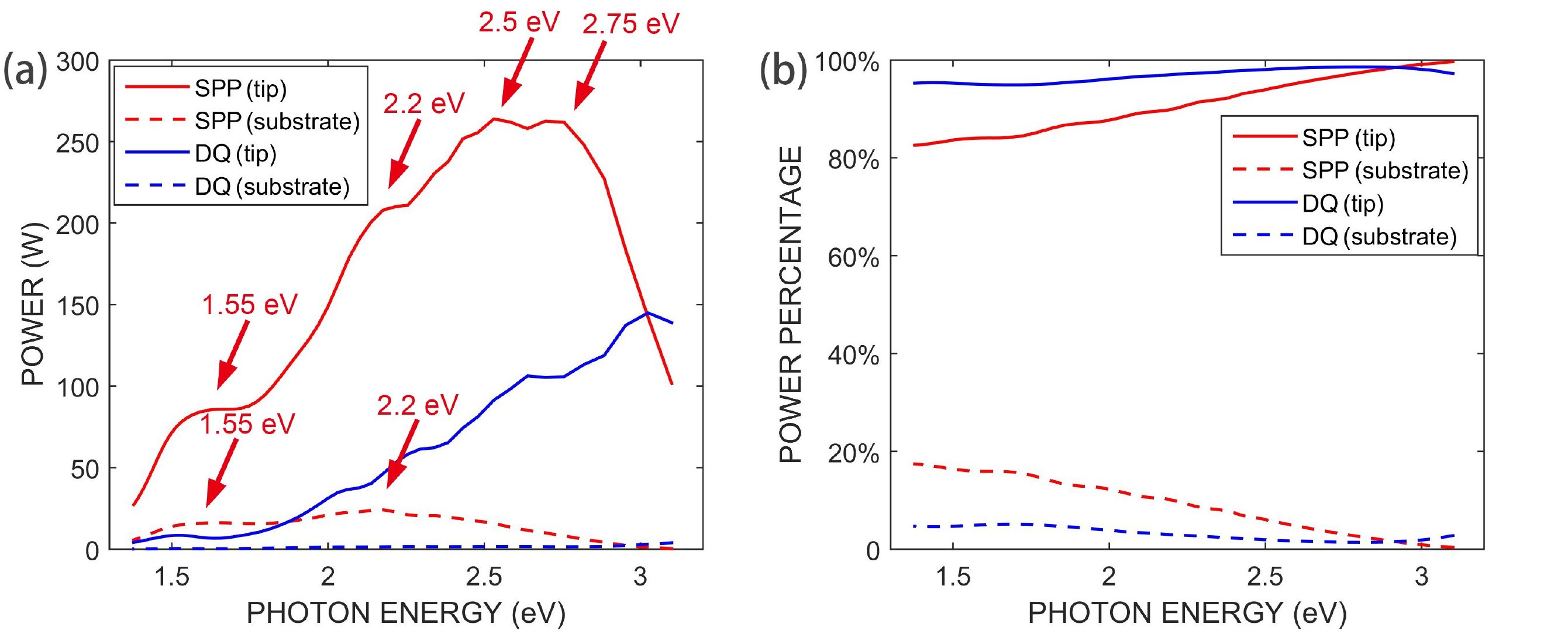}
\caption{ (a) Separate contribution to the SPPs (red) and direct quenching (blue) from tip (solid) and substrate (dashed), respectively. (b) Same as (a), but plotted in percentage.}
\label{fig3}
\end{figure}
We continue to discuss the different role of the metallic tip and substrate in the STML. The emission into SPP channels and  direct quenching occurs  at the apex of metallic tip and substrate. In aforementioned discussions, the SPPs of metallic tip and substrate  are considered as different channels, the direct quenching in these two places is considered as one channel. To analyze the difference of energy transfer property in tip and substrate, we present the fractional emission of SPP channel and direction quenching for both the tip and the substrate in \Fig{fig3}. In \Fig{fig3} (a), it shows  that the SPPs and the direct quenching  at the tip are much larger than that  at the surface of the substrate. The tip SPPs take the majority of the total power that excites the propagating SPP modes. The SPP propagating along the substrate has only two peaks, i.e., at 1.55 eV and 2.2 eV, in comparison with four peaks associated with the tip SPPs, the largest of which is at 2.5 eV. This shows that the tip SPP channel is dominant over the surface channels in the present setup. The direct quenching  at the tip is also dominant over that of the substrate, since the direct quenching  at the surface of the substrate is almost zero. Evidently, the typical shape of the tip, as well as the geometric structure between the tip and the substrate has substantially impact on the quenching process in the STML.

Figure 3 (b) shows the fractional contribution of the SPPs and direct quenching, distributed between the tip and the substrate. We find that the fraction of direct quenching into either tip or substrate  is approximately constant,  while the fraction of SPP emission channels into tip and substrate apparently changes monotonically as the frequency increases. Since the gap behaves as a plasmonic cavity in STML, the changing ratio indicates that the different surface shape will influence the SPP coupling characteristics at different frequencies. The almost constant quenching reveals the frequency independent characteristics of the quenching process in the STML. 

\section{Conclusion}
In summary, we studied the energy transfer into four different channels in the STML. Our main result is that, the majority of energy radiated by the tunneling electrons is transferred into SPP mode on the surface of the tip. The direct quenching in the apex of the tip also takes a large part. These two channels take most of the energy away, so that the energy that can be collected and utilized in the air mode and SPPs along the substrate surface is only a small portion of the total energy from inelastic electron tunneling. We propose possible methods that may increase the energy into the substrate SPPs with the help of our results. Firstly, the nano-structure with sharp apex geometry positioned on the substrate may help to increase the percentage of energy funneling into the substrate SPPs. Secondly, decreasing the non-radiative loss is important too, especially in the high frequency region. 

\section*{Appendix A: Electrical field inside the metallic region of the propagating  cylindrical SPPs along metal-air interface}
The electric field of SPP mode generated by an electric dipole emitter (located at $\bm r_0$ on z-axis) with the dipole moment being $\bm p$ can be given by $\bm E(\rho ,\varphi ,z) = i\omega_0 \mu_0 \bar{\bm G}_{SPP}(\bm r, \bm r_0) \bm p \delta(\bm r-\bm r_0)$, and the SPP contribution to the  Dyadic Green's function, i.e., $\bar{\bm G}_{SPP}(\bm r, \bm r')$, reads
\beq\label{Gspp}
\bar{\bm G}_{SPP}(\bm r, \bm r')=\int_{\rm{0}}^\infty \frac{{b}_{1}^2 {\kappa}_{  \rho}^{2 } e^{b_2{\kappa}_{\rho}(z+z^{'})}}{a(k_{SPP}^{2}-{\kappa}_{\rho}^{2})} d\kappa_\rho  \left( \begin{matrix}
 -J_{0}^{''}({{\kappa }_{\rho }}\rho )
& 0 
& -{{b}_{1}}J_{0}^{'}({{\kappa }_{\rho }}\rho ) 
\\0 
& -\frac{J_{0}^{'}({{\kappa }_{\rho }}\rho )}{{{\kappa }_{\rho }}\rho }
& 0
\\ {{b}_{1}}J_{0}^{'}({{\kappa }_{\rho }}\rho )
& 0 
& {{b}_{1}}^{2}{{J}_{0}}({{\kappa }_{\rho }}\rho ) \\
\end{matrix} \right),
\eeq 
where $\kappa_\rho$ is the amplitude of the in-plane wave vector, while $a=\pi{\sqrt{\epsilon _{1}(-\epsilon _{2})}} {\left(1-\frac{{\epsilon_1}^2}{{\epsilon_2}^2}\right)}
\frac{\epsilon _{1}+\epsilon _{2}}{\epsilon _{1}\epsilon _{2}} $, $ b_1=-\sqrt{{\epsilon _{1}}/{(-\epsilon _{2})}}$ and $ b_2=\sqrt{{(-\epsilon _{2})}/{\epsilon _{1}}}$, and  $\epsilon _{1}$ ($\epsilon _{2}$) is dielectric constant of air (metal). The $J_1$/$J_0$ is the first/zero order Bessel function, $k_{SPP}$ is the wave number of SPPs. For an electric dipole emitter (the magnitude of its dipole moment $\bm \mu$ is  1) orientated along z-axis, i.e., $\bm \mu=[0,0,1]^T$, the electric field  in the cylindrical coordinates can be calculated from \Eq{Gspp} as follows,
\beq\label{Efield}
\bm E(\rho ,\varphi ,z) =\int_{\rm{0}}^\infty 
\frac{{b}_{1}^3 {\kappa}_{  \rho}^{2 } e^{b_2{\kappa}_{\rho}(z+z^{'})}}{a(k_{SPP}^{2}-{\kappa}_{\rho}^{2})}
{d\kappa_{\rho}\left(
{{J}_{1}}({{\kappa }_{\rho }}\rho )
{{\bm{e}_{\rho }}}
+{{b}_{1}}{{J}_{0}}({{\kappa }_{\rho }}\rho )
{{\bm{e}_{z}}}\right)}.
\eeq

\begin{acknowledgments}
Y. Chen acknowledges financial support from the National Natural Science Foundation of China (Grant No. 61405067), and  the Fundamental Research Funds for the Central Universities, HUST: 2017KFYXJJ027. J.T. L\"u acknowledges financial support from the National Natural Science Foundation of China (Grant No. 61371015).
\end{acknowledgments}

%%%%%%%%%%%%%%%%%%%%%%% References %%%%%%%%%%%%%%%%%%%%%%%%%

\end{document}